\documentclass{emulateapj}

\newcommand{\cm}{\ensuremath{\rm cm}}
\newcommand{\days}{\ensuremath{\rm dy}}
\newcommand{\kms}{\ensuremath{\rm km\,s^{-1}}}

\newcommand{\mags}{\ensuremath{\rm mag}}

\newcommand{\kG}{\ensuremath{\rm kG}}

\newcommand{\ee}{\ensuremath{\rm e}}

\newcommand{\vpos}{\ensuremath{\mathbf{r}}}
\newcommand{\vposm}{\ensuremath{\tilde{\mathbf{r}}}}
\newcommand{\vdisp}{\ensuremath{\mathbf{a}}}

\newcommand{\mtrans}{\ensuremath{\mathbf{Q}}}

\newcommand{\Mstar}{\ensuremath{M_{\ast}}}
\newcommand{\Rpole}{\ensuremath{R_{\rm p}}}
\newcommand{\eratio}{\ensuremath{\varepsilon_{\ast}}}

\newcommand{\vsini}{\ensuremath{v \sin i}}

\newcommand{\srot}{\ensuremath{\Omega}}
\newcommand{\srotc}{\ensuremath{\srot_{\rm c}}}

\newcommand{\pot}{\ensuremath{\Phi}}

\newcommand{\intu}{\ensuremath{\mathcal{I}_{u}}}
\newcommand{\intl}{\ensuremath{\mathcal{I}_{\lambda}}}
\newcommand{\intz}{\ensuremath{\mathcal{I}_{\ast}}}

\newcommand{\tauu}{\ensuremath{\tau_{u}}}
\newcommand{\taul}{\ensuremath{\tau_{\lambda}}}

\newcommand{\source}{\ensuremath{\mathcal{S}}}

\newcommand{\dlambda}{\ensuremath{\Delta \lambda}}

\newcommand{\Halpha}{H$\alpha$}
\newcommand{\Hgamma}{H$\gamma$}
\newcommand{\Hdelta}{H$\delta$}

\newcommand{\sOriE}{$\sigma$~Ori~E}

\newcommand{\feros}{\textsc{feros}}

\begin{document}


\title{The rigidly rotating magnetosphere of $\sigma$ Ori E}
\shorttitle{The rigidly rotating magnetosphere of $\sigma$ Ori E}


\author{R. H. D. Townsend\altaffilmark{1}, 
        S. P. Owocki\altaffilmark{1} and 
        D. Groote\altaffilmark{2}}
\altaffiltext{1}{Bartol Research Institute, University of Delaware, Newark, DE 19716, USA; rhdt@bartol.udel.edu; owocki@bartol.udel.edu}
\altaffiltext{2}{Hamburger Sternwarte, Gojenbergsweg 112, 21029 Hamburg, Germany; dgroote@hs.uni-hamburg.de}
\shortauthors{Townsend, Owocki \& Groote}


\begin{abstract}
We attempt to characterize the observed variability of the magnetic
helium-strong star \sOriE\ in terms of a recently developed rigidly
rotating magnetosphere model. This model predicts the accumulation of
circumstellar plasma in two co-rotating clouds, situated in
magnetohydrostatic equilibrium at the intersection between magnetic
and rotational equators. We find that the model can reproduce well the
periodic modulations observed in the star's light curve, \Halpha\
emission-line profile, and longitudinal field strength, confirming
that it furnishes an essentially correct, quantitative description of
the star's magnetically controlled circumstellar environment.
\end{abstract}



\section{Introduction} \label{sec:intro}

As befits its status as the archetype of the helium-strong stars,
characterized by elevated photospheric helium abundances, the B2Vpe
star \sOriE\ is the most studied in its class. This is predominantly
due to the rich phenomenology of variability manifested by the star,
which exhibits modulations in its \Halpha\ Balmer emission
\citep{Wal1974}, its helium absorption-line strengths
\citep{PedTho1977}, its photometric indices \citep{Hes1976}, its UV
continuum flux and resonance-line strengths \citep{SmiGro2001}, its
6\,\cm\ radio emission \citep{LeoUma1993}, its linear polarization
\citep{KemHer1977}, and its longitudinal magnetic field strength
\citep{LanBor1978} --- all varying on the same 1.19\,\days\ period
identified with the stellar rotation cycle. Based on these extensive
observations, an empirical picture has emerged
\citep[e.g.,][]{GroHun1982,Bol1987,Sho1993} of the star as an
oblique-dipole magnetic rotator, whose field supports two
circumstellar clouds situated at the intersections between magnetic and
rotational equators.

Following on from previous attempts at devising a theoretical basis
for this picture \citep[see,
e.g.,][]{Nak1985,ShoBro1990,Bol1994,ShoBol1994}, \citet[][hereafter
TO]{TowOwo2005} have recently presented a new theoretical model for
the distribution of circumstellar plasma around stars, like \sOriE,
that are characterized both by strong magnetic fields (such that the
magnetic energy density is very much greater than the plasma energy
density) and by significant rotation. The model is built on the notion
that plasma, flowing along fixed trajectories prescribed by rigid
field lines, experiences an effective potential arising from a
combination of the stellar gravitational field and the centrifugal
force due to co-rotation. In the vicinity of points where the
effective potential undergoes a local minimum, as sampled along a
field line, the plasma can settle into a \emph{stable} configuration
of magnetohydrostatic equilibrium.

These concepts, originally advanced by \citet{MicStu1974} and
\citet{Nak1985}, are augmented in the TO treatment with a description
of the process filling the wells established by local potential
minima. In a manner similar to the \emph{magnetically confined wind
shock} (MCWS) model envisaged by \citet{BabMon1997a,BabMon1997b},
radiatively driven wind streams from opposing footpoints collide at
the top of closed magnetic loops, producing strong shocks. The
post-shock plasma cools via radiative emission, and then either falls
back to the star \citep[see, e.g.,][their Fig.~4]{udDOwo2002}, or ---
if the rotation is sufficient --- settles into an effective potential
well, where it continues to co-rotate in the circumstellar environment
and accumulate over time.

A strength of this \emph{rigidly rotating magnetosphere} (RRM) model
is its ability to predict quantitatively, via a semi-analytical
approach, the relative circumstellar plasma distribution for
\emph{arbitrary} magnetic field configurations --- not just for the
simple axisymmetric case of a rotation-axis aligned dipole, to which
magnetohydrodynamical simulations have so far been restricted on
grounds of computational tractability. In particular, application of
the RRM formalism to an oblique-dipole model star leads to a specific
prediction (cf. TO) of a density distribution that is sharply peaked
into two clouds, situated at the intersections between magnetic and
rotational equators. Such a distribution coincides with the
observationally inferred picture of \sOriE; thus, it is clear that the
RRM model holds promise for understanding both this particular star,
and related objects.

This letter presents results from a preliminary investigation into
whether the RRM model can \emph{simultaneously} reproduce the
spectroscopic, photometric and magnetic variability exhibited by
\sOriE. The following section discusses enhancements we have made to
the model since the TO study; Section~\ref{sec:model} then confronts the
predictions of the model with the observed behaviour of the star. We
discuss and summarize our findings in Section~\ref{sec:discuss}.


\section{Theoretical Developments} \label{sec:theory}

Although the RRM treatment presented by TO is quite general, much of
their analysis focuses on the specific case of a spherical star,
having a dipole magnetic field whose origin coincides with the stellar
origin. In the present analysis, we relax both of these
restrictions. Specifically, we allow the star to assume a
centrifugally distorted figure, by defining its surface not as a
sphere but as an isosurface of the effective potential function
$\pot(\vpos)$ (cf. TO, their eqn.~11). This isosurface is set at $\pot
= G \Mstar/ \Rpole$, where \Rpole\ denotes the polar radius of the
star, and --- here and throughout --- the other symbols have the same
meanings as in TO.

We also allow for the possibility of a decentered dipole field, by
introducing a vector \vdisp\ specifying the displacement of the
magnetic origin from the stellar origin. Then, the position vector in
the stellar reference frame, \vpos, is related to the magnetic-frame
position vector \vposm\ via
\begin{equation}
\vpos = \mtrans \vposm + \vdisp,
\end{equation}
where the orthogonal matrix \mtrans\ specifies a rotation by an angle
$\beta$ (the dipole obliquity) about the Cartesian $y$ axis. In the TO
treatment, \vdisp\ was set to zero, allowing simple expressions
(cf. their eqn.~21) to be obtained relating the spherical polar
coordinates in each reference frame. For non-zero \vdisp, such
expressions do not exist, but the transformation between frames
remains well defined.


\section{RRM Model} \label{sec:model}

\begin{deluxetable}{ccccc}
\tablewidth{0pt}
\tablecaption{RRM Model Parameters \label{tab:model}}
\tablehead{
  \colhead{\srot}    &
  \colhead{\eratio}  &
  \colhead{$\beta$}  &
  \colhead{$i$}      &
  \colhead{\vdisp}   \\
  \colhead{(\srotc)} &
  \colhead{}         &
  \colhead{(deg)}    &
  \colhead{(deg)}    &
  \colhead{(\Rpole)}
}
\startdata
0.5 & $10^{-3}$ & 55 & 75 & $(-0.041, 0.30, -0.029)$ \\
\enddata
\end{deluxetable}

\subsection{General Considerations} \label{ssec:general}

In Table~\ref{tab:model}, we summarize the basic parameters for the
RRM model of \sOriE. The rotation rate $\srot = 0.5\,\srotc$ is an
estimation based on the star's 1.19\,\days\ period, while the scale
height parameter \eratio\ (cf. TO, eqn.~38) has been assigned a value
typical to early-type stars. The choices of the obliquity, $\beta =
55\degr$, and observer inclination, $i = 75\degr$, have been guided by
certain characteristics of the spectroscopic and photometric
observations presented below. In particular, the unequal spacing of
the light-curve minima of \sOriE, whereby the secondary eclipse
follows the primary by $\sim 0.4$ of a rotation phase (see
Fig.~\ref{fig:photom}), restricts the possible RRM geometries to those
satisfying the approximate relation $\beta + i \approx 130\degr$. A
further indication that $i \gtrsim 75\degr$ comes from the fact that,
at lower inclinations, the \Halpha\ variations --- which take the form
of a double S-wave (see Fig.~\ref{fig:spectra}) --- would exhibit a
central spike when the two curves intersect; but no such spike is
observed in the spectroscopy.

The one remaining parameter of the model is the dipole offset vector
\vdisp. The motivation for assuming a non-zero offset comes from the
fact that the primary and secondary eclipses have unequal depths; and
that the two curves of the double S-wave are of unequal strength. The
offset we select comprises (i) a displacement of the dipole center by
$0.3\,\Rpole$ in a direction perpendicular to both magnetic and
rotation axes, followed by (ii) a displacement of the dipole center by
$-0.05\,\Rpole$ along the magnetic axis. The first displacement
produces a density contrast between the two clouds situated at the
intersections of the equators, that allows us to achieve a good fit
between the model and the observations. The second displacement has
little effect on the circumstellar plasma distribution, but its
introduction improves the fit to the magnetic field observations. We
note that \citet{ShoBol1994} also suggested an offset dipole field, in
order to explain asymmetries in the field observations.

\subsection{Photometry} \label{ssec:photom}

\begin{figure}
\epsscale{1.0}
\plotone{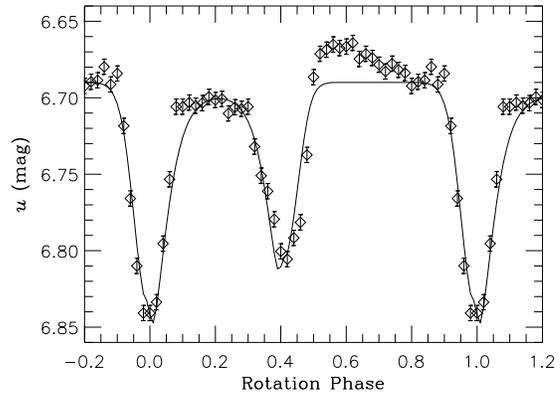}
\caption{Observed (diamonds) and modeled (solid line) Stromgren
$u$-band light curves of \sOriE, phased on the star's 1.19\,\days\
rotation period. The error bars on the observational data are based on
conservative estimates of the scatter in the raw data \citep[][their
Fig.~1]{Hes1977}.} \label{fig:photom}
\end{figure}

\begin{figure*}
\epsscale{1.0}          
\plotone{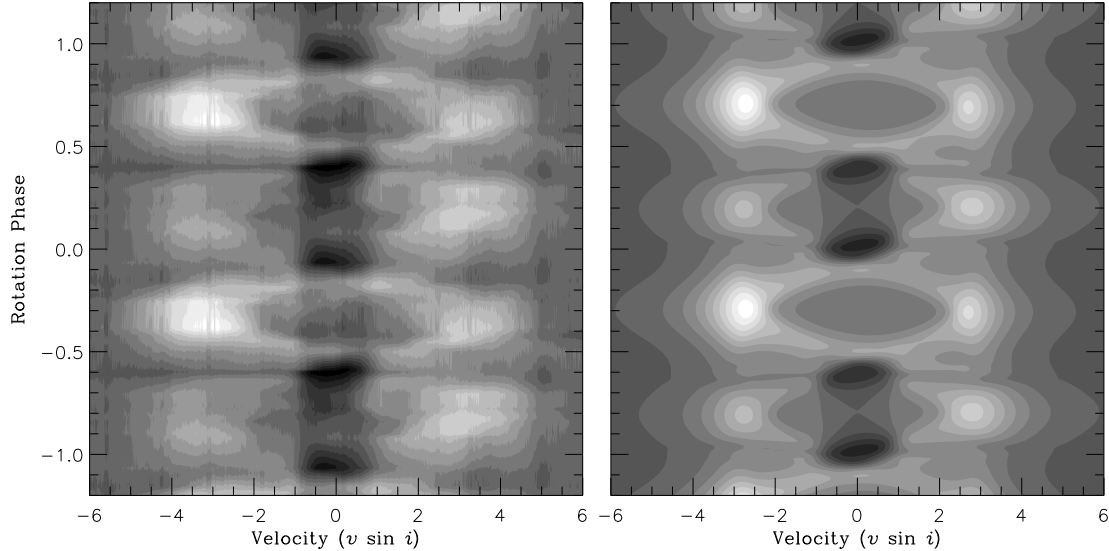}
\caption{Observed (left) and modeled (right) time-series grayscales of
the circumstellar \Halpha\ emission of \sOriE, phased on the star's
1.19\,\days\ rotation period. White indicates a 22\% excess (in
continuum units) over the background photospheric flux, and black
indicates a 12\% deficit; the grey levels are scaled linearly between
these extremes. The velocity axis is expressed in units of the star's
projected rotation velocity, $\vsini = 160\,\kms$ \citep{GroHun1982}.}
\label{fig:spectra}
\end{figure*}

In Fig.~\ref{fig:photom}, we show the observed Stromgren $u$-band
variations of \sOriE, obtained via electronic extraction from Fig.~2
of \citet{Hes1977}. Over these data we plot the synthetic light curve
predicted by the RRM model. To calculate this curve, we assume that
the photometric variations arise due to absorption of stellar flux by
intervening material confined within the co-rotating
magnetosphere. Then, the observed $u$-band specific intensity \intu,
along a given sightline, is evaluated via a simple attenuation
expression,
\begin{equation}
\intu = \intz\, \ee^{-\tauu},
\end{equation}
where \intz\ is the corresponding intensity at the stellar
photosphere, and \tauu\ is the $u$-band optical depth to the
photosphere. The variation of \intz\ across the stellar disk is
determined from a combination of the \citet{vonZei1924}
gravity-darkening law (with the ansatz that the intensity varies in
proportion to the bolometric flux) and the \citet{Edd1926}
limb-darkening law. We calculate \tauu\ from the density distribution
predicted by the RRM model, under the assumption that the opacity in
the magnetospheric material is spatially constant; this assumption
corresponds, for instance, to bound-free absorption processes in a
medium having uniform composition and excitation state.

Integrating \intu\ across the stellar disk, we arrive at the
synthetic light curve shown in Fig.~\ref{fig:photom}. Two free
parameters enter into this calculation. The normalization of \intz\
is chosen so that the intrinsic (un-obscured) $u$-band brightness of
the star is 6.69\,\mags. Likewise, the opacity and
density are together scaled to achieve a match with the observed depth
of the primary eclipse; this scaling corresponds to a maximal optical
depth, over all sightlines and rotation phases, of $\tauu = 1.63$.

The agreement between theory and observations is encouraging; the
synthetic light curve manages to capture the timing, duration and
strength of the eclipses. One obvious mismatch is that the RRM model
is unable to reproduce the emission feature at phase $\sim 0.6$,
during which the star appears to brighten above its intrinsic
level. However, based on the fact that this feature is conspicuously
\emph{absent} from the $u-b$ color variations \citep[cf.][]{Hes1977},
we believe the feature may arise from some as-yet-unknown photospheric
inhomogeneity, rather than from the circumstellar material. Further
monitoring is required to test this hypothesis.

\subsection{Spectroscopy} \label{ssec:spectra}

In Fig.~\ref{fig:spectra}, we plot the \emph{circumstellar} \Halpha\
variations exhibited by \sOriE, evaluated by subtracting a
phase-dependent synthetic absorption profile from spectroscopic
observations of the star (25 echelle spectra, obtained during
commissioning of the \feros\ instrument; see \citealp{Kau1999} and
\citealp{Rei2000}). The synthetic profiles are calculated using an
updated version of the model atmosphere code of \citet{Heb1983}, and a
photospheric model \citep[see][]{GroHun1997} that takes into account
the inhomogeneous helium surface distribution. The resulting synthetic
\Hdelta\ and \Hgamma\ profiles are in very good agreement with the
observations, on account of the absence of significant circumstellar
contamination in these lines. Assuming that the synthetic \Halpha\
profile is also representative of the true (but unknown) photospheric
profile, we may interpret the observed-minus-synthetic difference
profiles shown in Fig.~\ref{fig:spectra} as an estimated circumstellar
component of the total \Halpha\ flux.

Alongside the observational data, we plot the corresponding
predictions of the RRM model. We calculate the monochromatic specific
intensity \intl\ at wavelength $\lambda$, along a given sightline, via
formal solution to the equation of radiative transfer,
\begin{equation}
\intl = \intz\, \ee^{-\taul} + \source\,[1 - \ee^{-\taul}];
\end{equation}
here, \taul\ is the monochromatic optical depth, and \source\ is the
source function, assumed to be constant and wavelength independent
throughout the circumstellar environment.  We model the variation of
the photospheric intensity, across the stellar disk, using the same
function \intz\ adopted for the $u$-band synthetic photometry
(cf. Sec.~\ref{ssec:photom}); since we are ultimately interested only
in the circumstellar component of the \Halpha\ profile, we do not
include any photospheric profile. For sightlines that do not intersect
the disk, we set \intz\ to zero.

To derive the optical depth \taul, we assume an opacity proportional
to the local plasma density; this choice reflects the fact that
\Halpha\ absorption and emission, via radiative recombination, are
density-squared processes. We further assume that the wavelength
dependence of the opacity is a Gaussian of full width at half maximum
\dlambda, centered on the rest-frame wavelength, 6563\,\AA, of the
line. The modeled spectra are calculated by integrating \intl\ across
the stellar disk and surrounding circumstellar region. Three
parameters are involved in the synthesis, which we tune by hand to
achieve a good fit to the observations. This procedure leads to an
intrinsic line width $\dlambda = 30\,\kms$, a source function \source\
set at 0.4 of the maximal photospheric intensity, and an opacity and
density scaled such that the maximal optical depth achieved is $\taul
= 13.4$.

As with the photometry, there is good agreement between theory and
observations. The RRM model correctly reproduces all of the
qualitative features of the double S-wave \Halpha\ variations,
including the asymmetry between the red and blue wings, the differing
emission strengths at rotation phases $\sim 0.25$ and $\sim 0.75$, and
the absorption features at phases $\sim 0.0$ and $\sim 0.4$. The
phasing of the synthetic data is based on the photometric fit
(cf. Fig.~\ref{fig:photom}); thus, the small phase lag of the model,
relative to the observations, can wholly be attributed to the
uncertainties in the rotation period \citep[see][]{Rei2000} compounded
over the two decades separating the photometric and spectroscopic
datasets. The contrast between the absorption features seen in the
observations (strongest at phase $\sim 0.4$) and in the model
(strongest at phase $\sim 0.0$) arises, most likely, due to recent
changes in the star's surface abundance distribution \citep[see][his
Fig.~3]{Gro2003} that are not accounted for in our model.

There are, of course, some discrepancies; for instance, the maximal
separation of the S-wave curves is somewhat smaller in the model
($\sim 2.75\,\vsini$) than in the observations ($\sim
3.25\,\vsini$). Nevertheless, given the many simplifying assumptions
incorporated in the RRM model for \sOriE, it is quite striking how
well it matches the observed spectroscopic variability of this star.

\subsection{Magnetic Field} \label{ssec:magnetic}

\begin{figure}
\epsscale{1.0}
\plotone{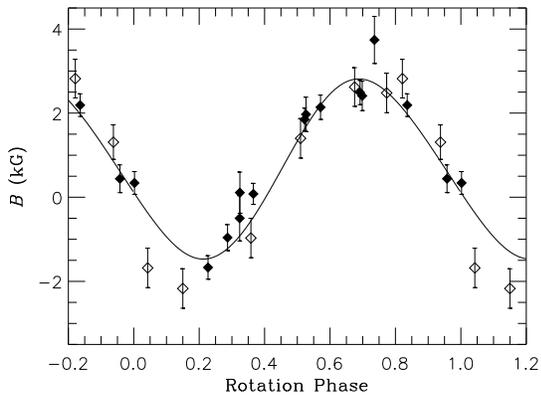}
\caption{Observed (diamonds) and modeled (solid line) longitudinal
magnetic field strength of \sOriE, phased on the star's 1.19\,\days\
rotation period. The open diamonds come from the \citet{LanBor1978}
dataset, while the filled diamonds are taken from the \citet{Boh1987}
dataset.}
\label{fig:field}
\end{figure}

In Fig.~\ref{fig:field}, we show the time-varying longitudinal
magnetic field strength of \sOriE, as measured by \citet{LanBor1978}
and \citet{Boh1987}; to maintain consistency with the spectroscopic
data, we rephase the field data using the same period and ephemeris
adopted by \citet{Rei2000}. Plotted over the observations is the
field strength predicted by the RRM model. These synthetic
data are calculated via an approach similar to \citet{Sti1950}: the
longitudinal (observer-directed) field is weighted by the local
photospheric intensity \intz\ introduced previously, integrated across
the stellar disk, and then renormalized by the disk-integrated
flux. The field geometry is specified by the parameters given in
Table~\ref{tab:model}, and we choose a field strength such that the
flux density is $11\,\kG$ at $\pm 1\,\Rpole$ along the magnetic
axis. Because the dipole is offset (cf. Sec.~\ref{ssec:general}), the
actual surface field strengths at the magnetic poles of the star are
somewhat different than this nominal value.

Once again, the agreement between theory and observations is
encouraging. The reduced chi-squared of the model, 1.79 for 20 degrees
of freedom, is rather larger than the value $\chi^{2}/n = 1.48$ found
by \citet{Boh1987} for a sinusoidal fit. However, these authors had
the luxury of adjusting the phase of their fit, whereas in the present
case the phasing is already constrained by the requirement that the
primary light minimum be at phase $0.0$
(cf.~Fig.~\ref{fig:photom}). We note that if the two outlier points at
phases $\sim 0.05$ and $\sim 0.15$ are discarded, then the
correspondence between model and observations improves
greatly. Whether these points are indeed erroneous is an matter that
should be resolved with further observations.


\section{Discussion \& Summary} \label{sec:discuss}

In the preceding sections, we demonstrate how the photometric,
spectroscopic and magnetic variability of \sOriE\ can be reproduced
extremely well by a rigidly rotating magnetosphere model. This is
persuasive evidence that the RRM paradigm furnishes an essentially
correct, \emph{quantitative} description of the star's magnetically
controlled circumstellar environment.

Future studies can now focus on refining the model and its input
parameters, beyond the exploratory treatment that we present
here. Furthermore, the analysis can be extended to other observable
quantities, such as radio emission, UV line strengths, and
polarization. Ultimately, based on our coarse exploration of parameter
space, we expect that a fully optimized RRM model will be able to
provide strong, independent constraints on the fundamental parameters
(e.g., mass, radius) of \sOriE, allowing fresh light to be shed on the
historical uncertainty \citep[see][]{Hun1989} concerning this
exceptional star's distance and evolutionary status.


\acknowledgments

This work is based in part on support by NASA grant LTSA04-0000-0060,
and NSF grant AST-0097983. We thank the \feros\ Consortium and ESO,
for access to the \sOriE\ spectroscopic data.


\bibliographystyle{aastex}
\bibliography{sigorie}

\begin{thebibliography}{}

\bibitem[\protect\citeauthoryear{{Babel} \& {Montmerle}}{{Babel} \&
  {Montmerle}}{1997a}]{BabMon1997a}
{Babel} J.,  {Montmerle} T.,  1997a, \apjl, 485, 29

\bibitem[\protect\citeauthoryear{{Babel} \& {Montmerle}}{{Babel} \&
  {Montmerle}}{1997b}]{BabMon1997b}
{Babel} J.,  {Montmerle} T.,  1997b, \aap, 323, 121

\bibitem[\protect\citeauthoryear{{Bohlender}, {Landstreet}, {Brown} \&
  {Thompson}}{{Bohlender} et~al.}{1987}]{Boh1987}
{Bohlender} D.~A.,  {Landstreet} J.~D.,  {Brown} D.~N.,    {Thompson} I.~B.,
  1987, \apj, 323, 325

\bibitem[\protect\citeauthoryear{{Bolton}}{{Bolton}}{1994}]{Bol1994}
{Bolton} C.~T.,  1994, \apss, 221, 95

\bibitem[\protect\citeauthoryear{{Bolton}, {Fullerton}, {Bohlender},
  {Landstreet} \& {Gies}}{{Bolton} et~al.}{1987}]{Bol1987}
{Bolton} C.~T.,  {Fullerton} A.~W.,  {Bohlender} D.,  {Landstreet} J.~D.,
  {Gies} D.~R.,  1987, in {Slettebak} A.,  {Snow} T.~P.,  eds, Proc. IAU
  Colloq. 92: Physics of Be Stars, p.~82

\bibitem[\protect\citeauthoryear{{Eddington}}{{Eddington}}{1926}]{Edd1926}
{Eddington} A.~S.,  1926, {The Internal Constitution of the Stars}.
Cambridge University Press, Cambridge

\bibitem[\protect\citeauthoryear{{Groote}}{{Groote}}{2003}]{Gro2003}
{Groote} D.,  2003, in {Balona} L.~A.,  {Henrichs} H.~F.,   {Medupe} R.,  eds,
  ASP Conf. Ser. 305: Magnetic Fields in O, B and A Stars: Origin and
  Connection to Pulsation, Rotation and Mass Loss, p.~243

\bibitem[\protect\citeauthoryear{{Groote} \& {Hunger}}{{Groote} \&
  {Hunger}}{1982}]{GroHun1982}
{Groote} D.,  {Hunger} K.,  1982, \aap, 116, 64

\bibitem[\protect\citeauthoryear{{Groote} \& {Hunger}}{{Groote} \&
  {Hunger}}{1997}]{GroHun1997}
{Groote} D.,  {Hunger} K.,  1997, \aap, 319, 250

\bibitem[\protect\citeauthoryear{{Heber}}{{Heber}}{1983}]{Heb1983}
{Heber} U.,  1983, \aap, 118, 39

\bibitem[\protect\citeauthoryear{{Hesser}, {Ugarte} \& {Moreno}}{{Hesser}
  et~al.}{1977}]{Hes1977}
{Hesser} J.~E.,  {Ugarte} P.~P.,    {Moreno} H.,  1977, \apjl, 216, 31

\bibitem[\protect\citeauthoryear{{Hesser}, {Walborn} \& {Ugarte}}{{Hesser}
  et~al.}{1976}]{Hes1976}
{Hesser} J.~E.,  {Walborn} N.~R.,    {Ugarte} P.~P.,  1976, \nat, 262, 116

\bibitem[\protect\citeauthoryear{{Hunger}, {Heber} \& {Groote}}{{Hunger}
  et~al.}{1989}]{Hun1989}
{Hunger} K.,  {Heber} U.,    {Groote} D.,  1989, \aap, 224, 57

\bibitem[\protect\citeauthoryear{{Kaufer}, {Stahl}, {Tubbesing}, {Norregaard},
  {Avila}, {Francois}, {Pasquini} \& {Pizzella}}{{Kaufer}
  et~al.}{1999}]{Kau1999}
{Kaufer} A.,  {Stahl} O.,  {Tubbesing} S.,  {Norregaard} P.,  {Avila} G.,
  {Francois} P.,  {Pasquini} L.,    {Pizzella} A.,  1999, The Messenger, 95, 8

\bibitem[\protect\citeauthoryear{{Kemp} \& {Herman}}{{Kemp} \&
  {Herman}}{1977}]{KemHer1977}
{Kemp} J.~C.,  {Herman} L.~C.,  1977, \apj, 218, 770

\bibitem[\protect\citeauthoryear{{Landstreet} \& {Borra}}{{Landstreet} \&
  {Borra}}{1978}]{LanBor1978}
{Landstreet} J.~D.,  {Borra} E.~F.,  1978, \apjl, 224, 5

\bibitem[\protect\citeauthoryear{{Leone} \& {Umana}}{{Leone} \&
  {Umana}}{1993}]{LeoUma1993}
{Leone} F.,  {Umana} G.,  1993, \aap, 268, 667

\bibitem[\protect\citeauthoryear{{Michel} \& {Sturrock}}{{Michel} \&
  {Sturrock}}{1974}]{MicStu1974}
{Michel} F.~C.,  {Sturrock} P.~A.,  1974, \planss, 22, 1501

\bibitem[\protect\citeauthoryear{{Nakajima}}{{Nakajima}}{1985}]{Nak1985}
{Nakajima} R.,  1985, \apss, 116, 285

\bibitem[\protect\citeauthoryear{{Pedersen} \& {Thomsen}}{{Pedersen} \&
  {Thomsen}}{1977}]{PedTho1977}
{Pedersen} H.,  {Thomsen} B.,  1977, \aaps, 30, 11

\bibitem[\protect\citeauthoryear{{Reiners}, {Stahl}, {Wolf}, {Kaufer} \&
  {Rivinius}}{{Reiners} et~al.}{2000}]{Rei2000}
{Reiners} A.,  {Stahl} O.,  {Wolf} B.,  {Kaufer} A.,    {Rivinius} T.,  2000,
  \aap, 363, 585

\bibitem[\protect\citeauthoryear{{Shore}}{{Shore}}{1993}]{Sho1993}
{Shore} S.~N.,  1993, in {Dworetsky} M.~M.,  {Castelli} F.,   {Faraggiana} R.,
  eds, Proc. IAU Colloq. 138: Peculiar versus Normal Phenomena in A-type and
  Related Stars, p.~528

\bibitem[\protect\citeauthoryear{{Shore} \& {Brown}}{{Shore} \&
  {Brown}}{1990}]{ShoBro1990}
{Shore} S.~N.,  {Brown} D.~N.,  1990, \apj, 365, 665

\bibitem[\protect\citeauthoryear{{Short} \& {Bolton}}{{Short} \&
  {Bolton}}{1994}]{ShoBol1994}
{Short} C.~I.,  {Bolton} C.~T.,  1994, in {Balona} L.~A.,  {Henrichs} H.~F.,
  {Contel} J.~M.,  eds, Proc. IAU Symp. 162: Pulsation; Rotation; and Mass Loss
  in Early-Type Stars, p.~171

\bibitem[\protect\citeauthoryear{{Smith} \& {Groote}}{{Smith} \&
  {Groote}}{2001}]{SmiGro2001}
{Smith} M.~A.,  {Groote} D.,  2001, \aap, 372, 208

\bibitem[\protect\citeauthoryear{{Stibbs}}{{Stibbs}}{1950}]{Sti1950}
{Stibbs} D.~W.~N.,  1950, \mnras, 110, 395

\bibitem[\protect\citeauthoryear{{Townsend} \& {Owocki}}{{Townsend} \&
  {Owocki}}{2005}]{TowOwo2005}
{Townsend} R.~H.~D.,  {Owocki} S.~P.,  2005, \mnras, 357, 251

\bibitem[\protect\citeauthoryear{{ud-Doula} \& {Owocki}}{{ud-Doula} \&
  {Owocki}}{2002}]{udDOwo2002}
{ud-Doula} A.,  {Owocki} S.~P.,  2002, \apj, 576, 413

\bibitem[\protect\citeauthoryear{{von Zeipel}}{{von Zeipel}}{1924}]{vonZei1924}
{von Zeipel} H.,  1924, \mnras, 84, 665

\bibitem[\protect\citeauthoryear{{Walborn}}{{Walborn}}{1974}]{Wal1974}
{Walborn} N.~R.,  1974, \apjl, 191, 95

\end{thebibliography}

\end{document}